\documentstyle[12pt]{article}
\newcommand{\m}{\medbreak}

\newcommand{\no}{\noindent}
\newcommand{\EQ}{\begin{equation}}
\newcommand{\eq}{\end{equation}}
\newcommand{\EQA}{\begin{eqnarray}}
\newcommand{\eqa}{\end{eqnarray}}
\newcommand{\ds}{\displaystyle}
\newcommand{\AR}{\renewcommand {\arraystretch}{1.5}
\begin{array}{l}}
\newcommand{\bAR}{\renewcommand {\arraystretch}{2}
\begin{array}{l}}
\newcommand{\ARc}{\renewcommand {\arraystretch}{1.5}
\begin{array}{c}}
\newcommand{\bARc}{\renewcommand {\arraystretch}{2}
\begin{array}{c}}
\newcommand{\ar}{\end{array} \renewcommand {\arraystretch}{1}}

\newcommand{\ee}{\mbox{$e^+e^-\ $}}

\newcommand{\PT}{\mbox{$p_T\ $}}
\newcommand{\ALL}{\mbox{$A_{LL}\ $}}
\newcommand{\ALLPV}{\mbox{$A_{LL}^{PV}\ $}}

\newcommand{\ALR}{\mbox{$A_{LR}\ $}}

\newcommand{\r}{\rightarrow}

\begin{document}
\begin{titlepage}
\vspace{0.2in}
\vspace*{1.5cm}
\begin{center}
{\large \bf Probing quark compositeness at hadronic colliders : \\
the case of polarized
beams \\}
\vspace*{0.8cm}
{\bf P. Taxil} and {\bf J.M. Virey}{$^1$}  \\ \vspace*{1cm}
Centre de Physique Th\'eorique$^{\ast}$, C.N.R.S. - Luminy,
Case 907\\
F-13288 Marseille Cedex 9, France\\ \vspace*{0.2cm}
and \\ \vspace*{0.2cm}
Universit\'e de Provence, Marseille, France\\
\vspace*{1.8cm}
{\bf Abstract \\}
\end{center}
A new handed interaction between subconstituents of quarks could be at
the origin of some small parity violating effects in one-jet
inclusive production.  Within a few years, the Relativistic Heavy Ion
Collider (RHIC)
will be used as a  polarized proton-proton collider. In this context, we
analyse the
possibilities of disentangling some new parity violating effects from the
standard spin
asymmetries which are expected due to the Standard Model QCD-Weak
interference. We also
explore the possibilities of placing some more stringent limits on the quark
compositeness scale $\Lambda$ thanks to measurements of such spin asymmetries.
\vspace*{1.0cm}

\vfill
\begin{flushleft}
\m\no
Number of figures : 4\\

\m\no
July 1995\\
CPT-95/P.3233\\
\m\no
anonymous ftp or gopher : cpt.univ-mrs.fr

------------------------------------\\
$^{\ast}$Unit\'e Propre de Recherche 7061

{$^1$} Moniteur CIES and allocataire MESR \\
email : Taxil@cpt.univ-mrs.fr

\end{flushleft}
\end{titlepage}

\section{Introduction}
\indent
\m
The idea of compositeness has been introduced in the hope of solving some of
the
problems left unanswered by the Standard Model (SM). In particular,
considering the
difficulties for explaining the family pattern of quarks and leptons, it
was natural to
speculate about the existence of an underlying substructure.
However, no "standard" model of quarks and leptons subconstituents (preons
or whatever)
has emerged so far. In phenomenological studies one
usually assumes that these subconstituents could interact by means of a
new "contact" interaction which is normalized to a certain
compositeness scale $\Lambda$. Then, one can describe the residual
interaction between quarks and/or leptons by 	using an effective lagrangian,
an approach which is valid at energies below the compositeness
scale.

The HERA $e-p$ collider is now deeply probing the presence of such an
anomalous term in
electron-quark scattering. However, hadronic colliders only would allow to
reveal
a new contact interaction belonging purely to the quark sector.
\m

Following \cite{EichtenEHLQ}, we
write the lowest dimensional current-current
interaction with a four fermion contact term under the form  :
\m
\EQ\label{Lcontact}
{\cal L}_{qqqq} = \epsilon \, {g^2\over {8 \Lambda_{qqqq}^2}}
\, \bar \Psi \gamma_\mu (1 - \eta \gamma_5) \Psi . \bar \Psi
\gamma^\mu (1 - \eta \gamma_5) \Psi
\eq
\noindent
where $\Psi$ is a quark doublet, $\epsilon$ is a sign and $\eta$ can take the
values $\pm 1$ or 0. $g$ is a new strong coupling constant
normalized usually to $g^2(\Lambda_{qqqq}) = 4\pi$.
\m
Working at Fermilab at the Tevatron ${\bar p} - p$ collider with
$\sqrt s = 1.8$ TeV, the CDF collaboration has published some bounds on the
scale
$\Lambda_{qqqq}$ (denoted shortly $\Lambda$ from now).
The strategy consisted in searching for an excess of events (compared to the
QCD
prediction at leading order) in the inclusive one-jet cross section
\cite{CDF1} and in the dijet invariant mass spectrum \cite{CDF2}. The
former gives the
best limit published today : $\Lambda > 1.4$ TeV \cite{PDG94}.
\m
The expression (eq. \ref{Lcontact}) is rather
general. In particular, there is no reason to assume that the new
interaction is a parity conserving (PC) one.  On the contrary, it has been
advocated for some time  \cite{EichtenEHLQ,AlbrightAdler} that parity
violation (PV)
could be present ($\eta \neq 0$).

It is then tempting to propose the search for an effect which is
absent in strong processes, like the production of jets,
as long as these processes are solely described in the framework of QCD
which is a
parity conserving theory.

\m
It is well-known, from deep-inelastic scattering (DIS) experiments and from
experiments at \ee colliders, that the measurement of some spin
asymmetries, either in
the final or in the initial state, gives a direct way to pin down a PV
interaction.

In the context of  hadronic colliders, the huge hadronic background makes
it very
difficult to measure the helicity state of a produced particle. Then, it is
mandatory to use polarized hadronic {\it beams} to build a spin
asymmetry.

Hadronic spin physics has been confined for a long time to fixed polarized
target
experiments but it is now under good way to reach the truly high energy domain.
Indeed,  stimulated by the
puzzling results obtained these last years from polarized DIS
experiments \cite{DISPolrecent},
the RHIC Spin Collaboration (RSC) \cite{RSC} has recently proposed
to run the Brookhaven Relativistic Heavy Ion Collider (RHIC) in the $pp$ mode,
with longitudinally (or transversely) polarized beams. The degree of
polarization of the beam will be as high as 70\%, with a high luminosity
at a center-of-mass energy up to 500 GeV. This proposal have been approved
recently and
a very complete program of measurements of (PV or PC) spin asymmetries will be
performed (see ref. \cite{BuncePw} ; for reviews on spin physics at future
hadronic
colliders one can consult refs. \cite{BRST,Rivista} ).

In a first step, the RSC will focus on the PC double
spin asymmetry \ALL in jet production, direct photon production and other
hadronic
processes governed by QCD.  For an inclusive process like $p_a\ p_b \ \r c
+ X$, where
$c$ is either a jet or a well-defined particle, \ALL is defined as (in obvious
notations, the signs  $\pm$ refer to the helicities of the colliding protons) :
\EQ
\label{ALLdef}
A_{LL} ={d\sigma_{a(+)b(+)}-d\sigma_{a(+)b(-)}\over
d\sigma_{a(+)b(+)}+d\sigma_{a(+)b(-)}}
\eq

\m
On the other hand, large Standard PV effects should be obtained in
the direct production of the $W$ and $Z$ gauge bosons
\cite{BRST,BouGuiSof}. To build a PV asymmetry
one single polarized beam is sufficient
$A_{LR} =
 ({d\sigma_{a(-)b}-d\sigma_{a(+)b})/
(d\sigma_{a(-)b}+d\sigma_{a(+)b}}) $.

One can also define a double helicity PV asymmetry :
\EQ
\label{ALLPVdef}
A_{LL}^{PV} ={d\sigma_{a(-)b(-)}-d\sigma_{a(+)b(+)}\over
d\sigma_{a(-)b(-)}+d\sigma_{a(+)b(+)}}
\eq
\noindent

\m
A whole set of measurements of the large
Standard PC and PV asymmetries should allow to isolate with very good
precision the
polarized distribution functions of the various partons (quarks and gluons) in
a
polarized proton (see refs. \cite{BouGuiSof,DoncheskiHalzen,BouSof1}) and,
in the
meantime, to perform some polarization tests of the Standard Model. Note that
transversely polarized beams, whose great interest has been emphasized
\cite{BuncePw} will be also available at RHIC.

\m
In this letter we focus on the production of a single jet in polarized $pp$
collisions at RHIC with $\sqrt{s} \ =\ 500$ GeV. From now d$\sigma$ means

\EQ
\label{dsrap0}
d\sigma \; \equiv \; {d^2\sigma \over {dp_T dy}}\; \; {\rm at} \; y\, =\, 0
\eq
\no
where $p_T$ is the jet transverse momentum and $y$ its rapidity.
\m
We will concentrate on a high $p_T$ region where quark-quark elastic
scattering is the dominant process and where an effect due to compositeness has
some chance to be observed. At RHIC it corresponds to the region
$60\, < p_T <\, 120$ GeV/c.
\m
The interest of using  a PV spin asymmetry to reveal a new effective handed
interaction between quarks has been already noticed in the literature
\cite{BRST,BouGuiSof,TannenbaumPenn}. It is however valuable
to explore this issue in more details.
\m
First, QCD-Electroweak boson exchange interference terms
\cite{AbudBaurGloverMartin,PaigeRanft} which
are indeed present, not only in the vicinity of the W and Z "Jacobian
peaks", but
also at high $p_T$, have been essentially neglected up to now
\cite{BRST,BouGuiSof}. It is mandatory to take
these terms into account to set definitive conclusions about the origin of
any PV
effect which should be observed in this context. The calculations described
below take
into account all the  (lowest order) relevant terms : QCD + Electroweak(EW)
+ Contact
Terms (CT) which are to be added coherently.
\m
The second point is more technical : various choices of polarized
distribution functions of quarks and antiquarks in a polarized proton are now
available, in particular those which have been updated according to the latest
polarized DIS results. Also, at a variance with previous works, we will
concentrate on
the asymmetry \ALLPV from which the largest PV effects should be seen.
\m

\section{Parity violating subprocesses for the one-jet inclusive production}
\indent
\m
The expression of $d\sigma$, eq.(\ref{dsrap0}) is given by the well-known
formula:
\EQ
\label{sigjet}
\AR\ds
d\sigma \; =\;
\sum_{ij}{2\, p_T\over1+\delta_{ij}}\int_{x_{min}} ^{1} dx_a
 \left( x_ax_b \over x_a - p_T/\sqrt s  \right)
\biggl[f^{(a)}_i\bigl(x_a,Q^2\bigl) f^{(b)}_j\bigl(x_b,Q^2\bigl)
{d\hat \sigma(i,j) \over d\hat t} \left( \hat s, \hat t, \hat u
\right)
\\
\hskip 8truecm + \;(i\leftrightarrow j) \biggl]
\ar \eq
\no
where the sum runs over all the various partons.
\m
For consistency, we will follow the CDF analysis,
restricting to leading order terms. For the scale $Q^2$, we have taken $Q^2
\, =\,
p_T^2$ after having checked that changing this value between $p_T^2/4$ and
$4.p_T^2$
has a very small influence on our results on \ALLPV.

\m
The helicity dependent cross section is given by ($h_{a,b}$ refers to the
helicities
of the protons and $\lambda_{1,2}$ to those of the partons) :
\EQ\label{hcrosssection}
\AR\ds
d\sigma^{h_a,h_b}\; =\;
\sum_{ij}{2\, p_T\over 1+\delta_{ij}}\sum_{1,2}\int_{x_{min}} ^{1} dx_a
 \left( x_ax_b \over x_a - p_T/\sqrt s  \right)
\biggl[f^{(h_a)}_{i,\lambda_1}\bigl(x_a\bigl)
       f^{(h_b)}_{j,\lambda_2}\bigl(x_b\bigl)
{{d\hat \sigma}^{\lambda_1,\lambda_2}\over d\hat t} (i,j)
\\
\hskip 8truecm + \;(i\leftrightarrow j) \biggl]
\ar\eq
\no
where we have dropped out the scale dependance for simplicity.
Following the notations of ref.\cite{BouGuiSof} we have :
\EQ
{{d\hat \sigma}^{\lambda_1,\lambda_2}\over d\hat t}(i,j)
\; =\;{\pi\over \hat s^2} \,
\sum_{\alpha,\beta} T_{\alpha,\beta}^{\lambda_1,\lambda_2}(i,j)
\eq
\no
$T_{\alpha,\beta}^{\lambda_1,\lambda_2}(i,j)$ denoting the matrix element
squared with
$\alpha$ boson and $\beta$ boson exchanges, or with one exchange process
replaced by a
contact interaction.
\m

If one ignores the contribution of the antiquarks, which is marginal in our \PT
range, and restricts to the main channels $q_iq_i \r q_iq_i$
and $q_iq_j \r q_iq_j$ ($i\neq j$) one gets in short :
\EQ
\label{ALLPVjet}
A_{LL}^{PV}\ .\ d\sigma \simeq \; \sum_{ij} \sum_{\alpha,\beta}
\int
\left(T_{\alpha,\beta}^{--}(i,j) - T_{\alpha,\beta}^{++}(i,j)
\right)
\biggl[q_i(x_a)\Delta q_j(x_b) + \Delta q_i(x_a)q_j(x_b)
+ \;(i\leftrightarrow j) \biggl]
\eq
\noindent
where we have introduced as usual
the polarized quark distributions : \linebreak
$\Delta q_i(x,Q^2)\ =\ q_{i+} - q_{i-}$,
$\, q_{i\pm}(x,Q^2)$ being the distributions of
the polarized quark of flavor $i$, either
with helicity parallel (+) or antiparallel (-) to the parent proton
helicity. In eq.(\ref{ALLPVjet}),
$d\sigma$ is given by eq.(\ref{sigjet}).  Concerning the QCD
contribution to $d\sigma$, we take also into account the antiquarks an also
$q(\bar q)g$ and $gg$ scattering although these subprocesses are not
dominant in the
high \PT region we consider.

\m

\subsection{Standard QCD-Electroweak interference effects}
Concerning the influence of QCD-EW interference terms
on \ALR or \ALLPV, in the high  $p_T$ regime we are
concerned with, only some estimates can be found in the literature
\cite{TannenbaumPenn} since previous authors focused mainly on the  $p_T \sim
M_{W,Z}/2$ region which is dominated by the s-channel $W$ and $Z$ resonance
contributions \cite{PaigeRanft}.
\m
We give in Fig.1 the asymmetry \ALLPV in one-jet production at RHIC which
is expected
from purely QCD-EW interference terms. The correct expressions for the
$T_{\alpha,\beta}$'s can be found in ref. \cite{BouGuiSof}. It can be
checked that
90\% of the effect comes both from the interference terms $T_{gZ}$ between
the gluon and Z exchange graphs (identical quarks) and from the terms
$T_{gW}$ (quarks
of different flavors).
\m
We have used various sets of polarized distributions, some quite old, like BRST
\cite{BRST}, CN1 \cite{CN1} or CN2 \cite{CN2}, and some recent ones which
give better
fits to the new polarized DIS data : BS \cite{BS95} and GS.a,b,c labelled
according to
ref. \cite{GS}.
Note that when some distributions, like GSa or b or c, differ by the shape
of the
polarized gluonic contribution, they give essentially the same values for
\ALLPV.
Since BS and GS distributions provide two extreme cases we will keep only
these latter
in the following.

The rise of \ALLPV with \PT is due to the increasing importance of quark-quark
scattering relatively to other terms involving gluons. \ALLPV remains small
(at most
4\% at $p_T=100$ GeV/c) but it is measurable with the sensitivity available
at RHIC
(see below).

\subsection{Interference between Contact and Standard amplitudes}
One has to consider (schematically) all the terms appearing in
$\mid QCD + EW + CT \mid^2$. Since we restrict to values of $\Lambda$ above
the CDF bound, the squared terms $\mid CT \mid^2$ involving  $1/\Lambda^4$ are
negligible at RHIC except for some unreasonable values of \PT.
Any effect should come from CT+Standard interference terms.

One can find in ref. \cite{Rivista} all the helicity dependent $\mid QCD +
CT\mid^2$
terms for the scattering of all kinds of quarks and antiquarks.
As long as quarks are concerned, only identical quarks give a contact amplitude
interfering with the one gluon exchange amplitude:
\EQ\label{gCT}
T_{g.CT}^{\lambda_1,\lambda_2}(i,i) \; =\;
{8 \over 9} \alpha_s\, {\epsilon \over \Lambda^2} \,
(1 - \eta \lambda_1)(1 -  \eta\lambda_2)\left({\hat s^2 \over \hat t} +
{\hat s^2 \over \hat u}\right)
\eq
\m
On the other hand, we have found that the interference terms between
Electroweak and
Contact amplitudes cannot be ignored although they are not the source of
the main
effect. In this case, identical quarks as well as quarks of different
flavors are
involved.
We have :
\m \no
- for $q_iq_i \r q_iq_i$ :
\EQ \label{qiqiZ}
T_{Z.CT}^{\lambda_1,\lambda_2}(i,i) \, =\, {4\, \alpha_Z \over 3}
{\epsilon \over \Lambda^2} \,
\biggl[ (1 - \lambda_1)(1 - \lambda_2) C_L^2 (1+\eta)
+ (1 + \lambda_1)(1 + \lambda_2) C_R^2 (1-\eta)\biggl]
\left({\hat s^2 \over \hat t_Z} +
{\hat s^2 \over \hat u_Z}\right)
\eq \no
where $\alpha_Z = \alpha/\sin^2\theta_W\cos^2\theta_W$ ,
$C_{R[L]} = -e_i\sin^2\theta_W\, [T^3_{q_i}-e_i\sin^2\theta_W]$
for a quark of charge $e_i$ and $\hat t_Z[\hat u_Z] = \hat t[\hat u] - M^2_Z$.
\no
$T_{\gamma .CT}^{\lambda_1,\lambda_2}(i,i)$ can be obtained from
eq.(\ref{qiqiZ})
by changing
$\, \alpha_Z
\r \alpha.e_i^2$, $C_{L,R} \r 1$ and $\hat t_Z[\hat u_Z] \r \hat t[\hat u]$.
\m \no
- for $q_iq_j \r q_iq_j$ $i \neq j$:
\EQ\label{qiqjZ}
T_{Z.CT}^{\lambda_1,\lambda_2}(i,j) \; =\;
 \alpha_Z
{\epsilon \over \Lambda^2} \,
\biggl[ (1 - \lambda_1)(1 - \lambda_2) C_L^i C_L^j(1+\eta)
+ (1 + \lambda_1)(1 + \lambda_2) C_R^i C_R^j (1-\eta)\biggl]
{\hat s^2 \over \hat t_Z}
\eq \no
$T_{\gamma .CT}^{\lambda_1,\lambda_2}(i,j)$ being obtained from
eq.(\ref{qiqjZ})
by changing $\, \alpha_Z \r \alpha e_ie_j$ , $\forall C \r 1$ and $\hat t_Z
\r \hat t$,
\m\no
and also
\EQ
T_{W.CT}^{\lambda_1,\lambda_2}(i,j) \; =\;
{ \alpha_W \over 6}
{\epsilon \over \Lambda^2} \, \mid V_{ij}^{CKM} \mid^2
\biggl[ (1 - \lambda_1)(1 - \lambda_2) (1+\eta)
\biggl] {\hat s^2 \over \hat u_W}
\eq \no
where $\alpha_W = \alpha/\sin^2\theta_W$  and $\hat u_W = \hat u - M^2_W$.
\m
In the actual calculations we have added all the terms, involving quarks or
antiquarks,
dominant or not.

\section{Discussion and results}

\m
At RHIC, a very important parameter is the high luminosity which increases with
the energy, reaching ${\cal L} = 2. 10^{32} \, cm^{-2}.s^{-1}$ at $\sqrt s
= 500$ GeV.
These figures yield an integrated luminosity
$L_1\ = \ \int{\cal L} dt $ = 800 $pb^{-1}$ in a few
months running. In the following, we will call $L_2$ the luminosity giving
four times
this sample of events. We have integrated over a \PT bin of 10 GeV/c which
is typical
of a CDF like detector \cite{CDF1} in this \PT range.

This high luminosity will allow some very small
statistical uncertainties on a spin asymmetry like \ALLPV \cite{BuncePw}.
This uncertainty is given by :
\EQ \label{error}
\Delta A\; =\; {1\over {\cal P}^2}\, {2\over (N_{++}+N_{--})^2}
{\sqrt{N_{++}N_{--}
(N_{++}+N_{--})}}
\eq
\no
where $N_{++}$($N_{--}$) is the expected number of events in the helicity
configuration $++$ ($--$) and $\cal P\, =\, $ 0.7  is the degree of
polarization
of one beam. $\Delta A$ is roughly equal to $2/ \sqrt {(N_{++}+N_{--})}$.
One gets $\Delta A = \pm 0.01 (0.005)$ with $N_{++}+N_{--} \sim 40000 (170000)$
events. These figures are not unrealistic, even at high \PT, thanks to the high
integrated luminosities $L_1$ or $L_2$.
\m
We present in Fig. 2 the results of our complete calculation for \ALLPV,
including
all terms, for $\Lambda = 1.4\,$ TeV (that is the present CDF bound).
The expected Standard asymmetry is shown for comparison. The parameter
governing the sign of \ALLPV is the sign of the product $\epsilon .\eta$.
As can be seen from eq.(\ref{gCT}), which gives the main effect,
since $\hat t$ and $\hat u$ are negative, $\epsilon = -1\, (+1)$ corresponds
to constructive (destructive) interference. We have chosen the BS
parametrization for
illustration. One can see that, at RHIC, even with the integrated
luminosity $L_1$,
it is very easy to separate the Standard from the Non-Standard cases.
With GS distributions, the magnitudes of the asymmetries are reduced but
the effect
is still spectacular.
\m
For $\Lambda = 2$ TeV (Fig. 3), with $L_2$ there is still a 4$\sigma$
difference
(2$\sigma$ with $L_1$)  between the Standard and Non-Standard asymmetries,
especially at values of \PT above 80 GeV/c. The small positive value of \ALLPV
for $\epsilon .\eta = +1$ at low \PT is due to the influence of the
QCD-EW and CT-EW interference terms.
\m
In Fig. 4 we display again \ALLPV with $\Lambda = 2$ TeV, but now
calculated using the
two extreme choices, BS and GS distributions. The clearest result is that,
in spite of
the present uncertainty due to the imperfect knowledge of the polarized quark
distributions, a value for \ALLPV close to zero at large \PT is the sign of the
presence of Non-Standard physics (namely either a left-handed contact
interaction with
destructive interference or a right-handed one with constructive interference).
Indeed, with $L_2$, the two close BS and GS curves stand at $3\sigma$ from
the smaller
QCD-EW asymmetry which corresponds to the GS parametrization. The situation
is less
spectacular in the case where $\epsilon .\eta = -1$ but it is still
interesting.
\m

Finally, we have tried to determine if the information from the measurement
of \ALLPV
could compete with the bounds on $\Lambda$ one could reach in the future at the
Tevatron (with unpolarized beams). Following the stategy of refs.
\cite{EichtenEHLQ,ChiapPerLHC},  we have calculated $d\sigma$(QCD+CT) in
$p\bar p$
collisions at $\sqrt s = 1.8$ TeV using MRS \cite{MRS} distributions and
demanding a
100\% deviation from the QCD prediction at large \PT (with at least 10
QCD events). Our crude estimate gives amazingly exactly the same result as the
published sophisticated CDF study ($\Lambda > 1.4\,$TeV with 4.2 $pb^{-1}$ of
integrated luminosity). With an integrated luminosity of 100 $pb^{-1}$ we
expect a
limit of  $\Lambda > 2$ TeV at the Tevatron. We give in Table 1 the 95\%
C.L. limits
on $\Lambda$ which should be obtained at RHIC from the measurement of \ALLPV.
\m
\begin{table}[t]
\begin{center}
\begin{tabular}{|c|c|c|}
\hline
\bf $\Lambda$ & $\epsilon .\eta = -1$ & $\epsilon .\eta = +1$\\
\hline
$L_1$ & BS : 2200 & 2070\\  \cline{2-3}
 & GS : 1950 & 1900 \\
\hline
$L_2$ & BS : 3050 & 3010\\  \cline{2-3}
 & GS : 2710 & 2670 \\
\hline
\end{tabular}
\end{center}
\caption{
Limits on $\Lambda$ (in GeV) from the measurement of \ALLPV at RHIC with the
integrated
luminosities $L_1$ and $L_2$, according to BS and GS polarized distributions.}
\end{table}
\m
\section{Conclusion}
It has been stressed for some time that polarization at
hadronic colliders should improve their potential capabilities
\cite{BRST,Rivista}, in
particular in the search for New Physics if the energy is as large as the
LHC energy
(see e.g. \cite{FTcasal}).
We have seen here that, in spite of the lower energy, the RHIC collider,
running in
the $pp$ mode, could compete with the Tevatron, thanks to the polarization
and also to
the high luminosity which turned out to be the key factor for our analysis.
The high statistics which should be available could allow to disentangle a
Non-Standard \ALLPV due to compositeness from the Standard asymmetry due to
QCD-EW
interference, provided $\Lambda$ lies in the  2 - 3 TeV range. Needless to
recall that,
if such a signal is observed, then a unique information could be obtained on
the
chirality structure of the new contact interaction from the determination
of the sign
of the product $\epsilon .\eta$.

It is true that the present imperfect knowledge of the polarized quark
distributions still induces some uncertainties. It has to be stressed
however that our
wisdom on this subject will change drastically in the near future, thanks to
new
polarized DIS experiments (the experiment HERMES at HERA
\cite{HERMES} is presently running) and to the RHIC Spin Collaboration
program itself.

\vspace*{3cm}

\no {\bf Acknowledgments}

\m

We are indebted to C. Benchouk,  C. Bourrely, P. Chiappetta,
M.C. Cousinou, A. Fiandrino, M. Perrottet
and J. Soffer for discussions, help and comments, and to T. Gehrmann and
W.J. Stirling
for providing us some computer program about the GS distributions. Thanks are
also
due to E. Kajsfasz for information on the CDF
results.

\m


%


\newpage  \no
{\bf Figure captions}
\bigbreak
\no
{\bf Fig. 1} \ALLPV for one-jet inclusive production from QCD-EW interference
only, versus $p_T$, with RHIC parameters (see Section 3) : $pp$ collisions,
$\sqrt s = 500$ GeV, integrated luminosities $L_1$ (large error bars) and $L_2$
(small
error bars), according to various choices of polarized distributions : BS
(plain
curve), BRST (dotted), CN1 and CN2 (dot-dashed), GSa,b,c (dashed curves).
\bigbreak
\no
{\bf Fig. 2} \ALLPV versus \PT for $\Lambda = 1.4\,$ TeV.
$\epsilon . \eta = -1$ (dashed curve) ; $\epsilon . \eta = +1$ (dot-dashed
curve),
and pure QCD-EW interference (plain curve). The calculations are
performed with BS distributions and the RHIC parameters are the same as in
Fig. 1
\bigbreak
\no
{\bf Fig. 3} Same as Fig. 2 for $\Lambda = 2\,$TeV.
\bigbreak
\no
{\bf Fig. 4} \ALLPV  versus \PT at RHIC with $\Lambda = 2\,$ TeV
for BS (plain curves) and GS (dot-dashed curves) polarized distributions.
The error bars correspond to the luminosity $L_2$.
\bigbreak

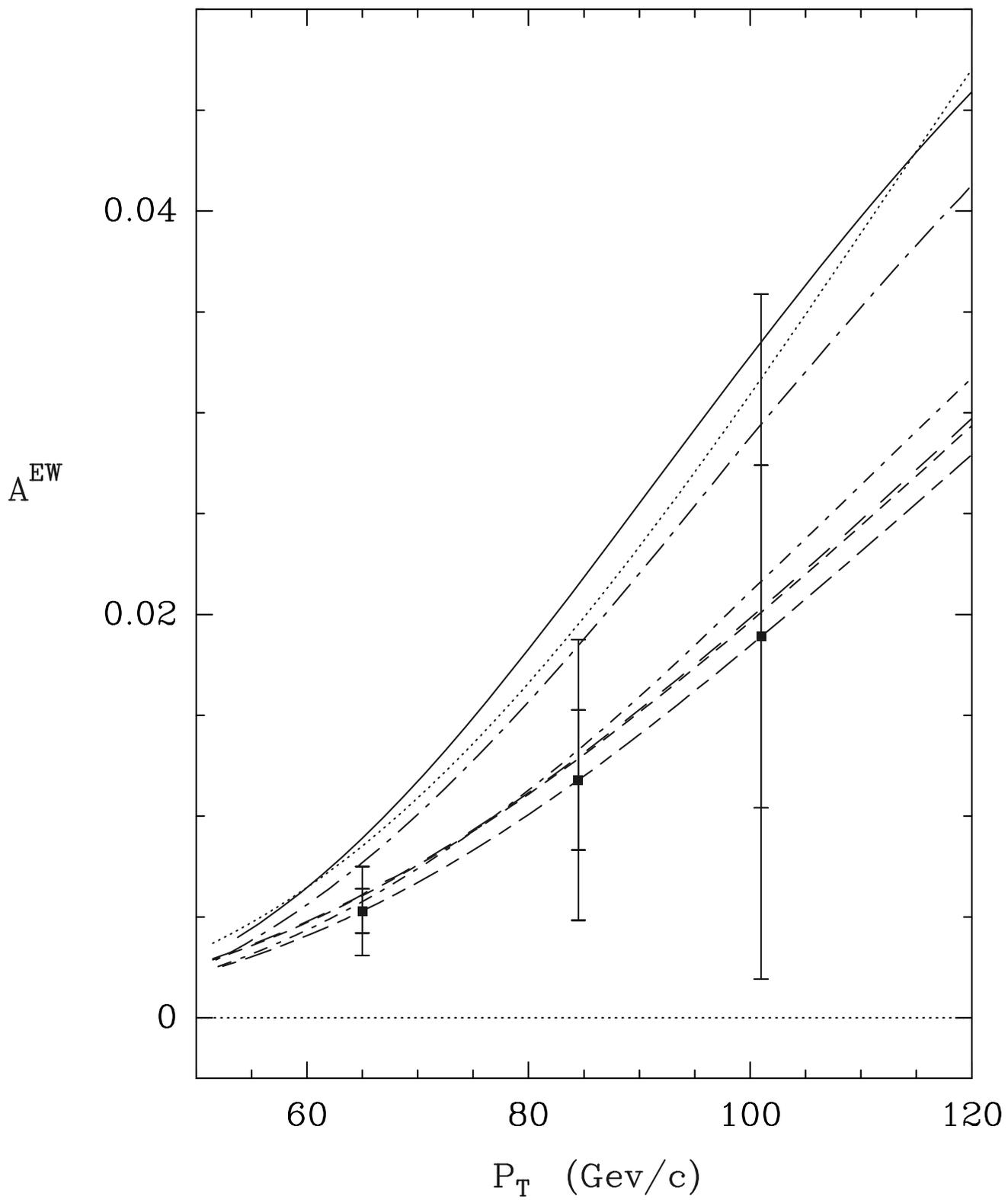
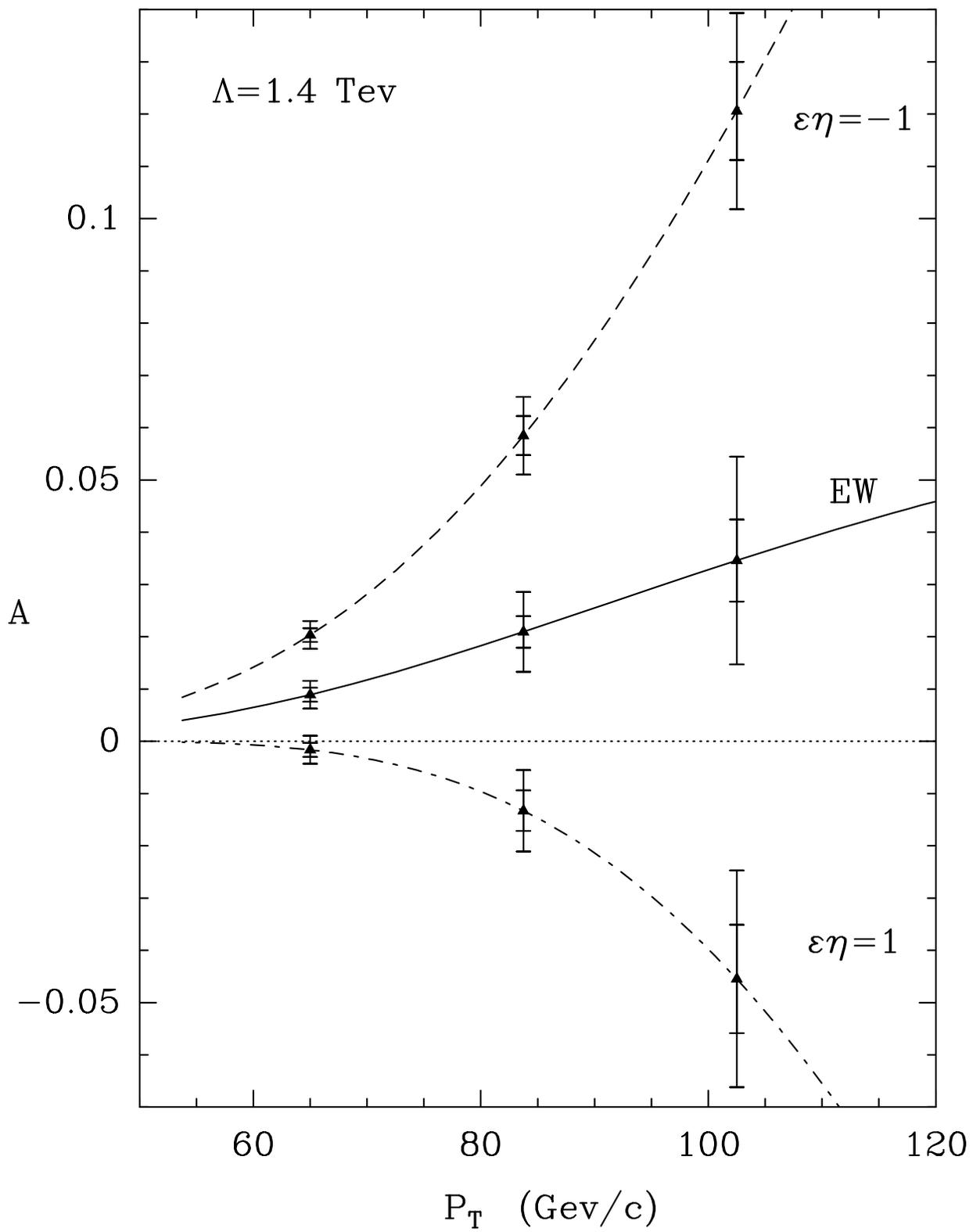
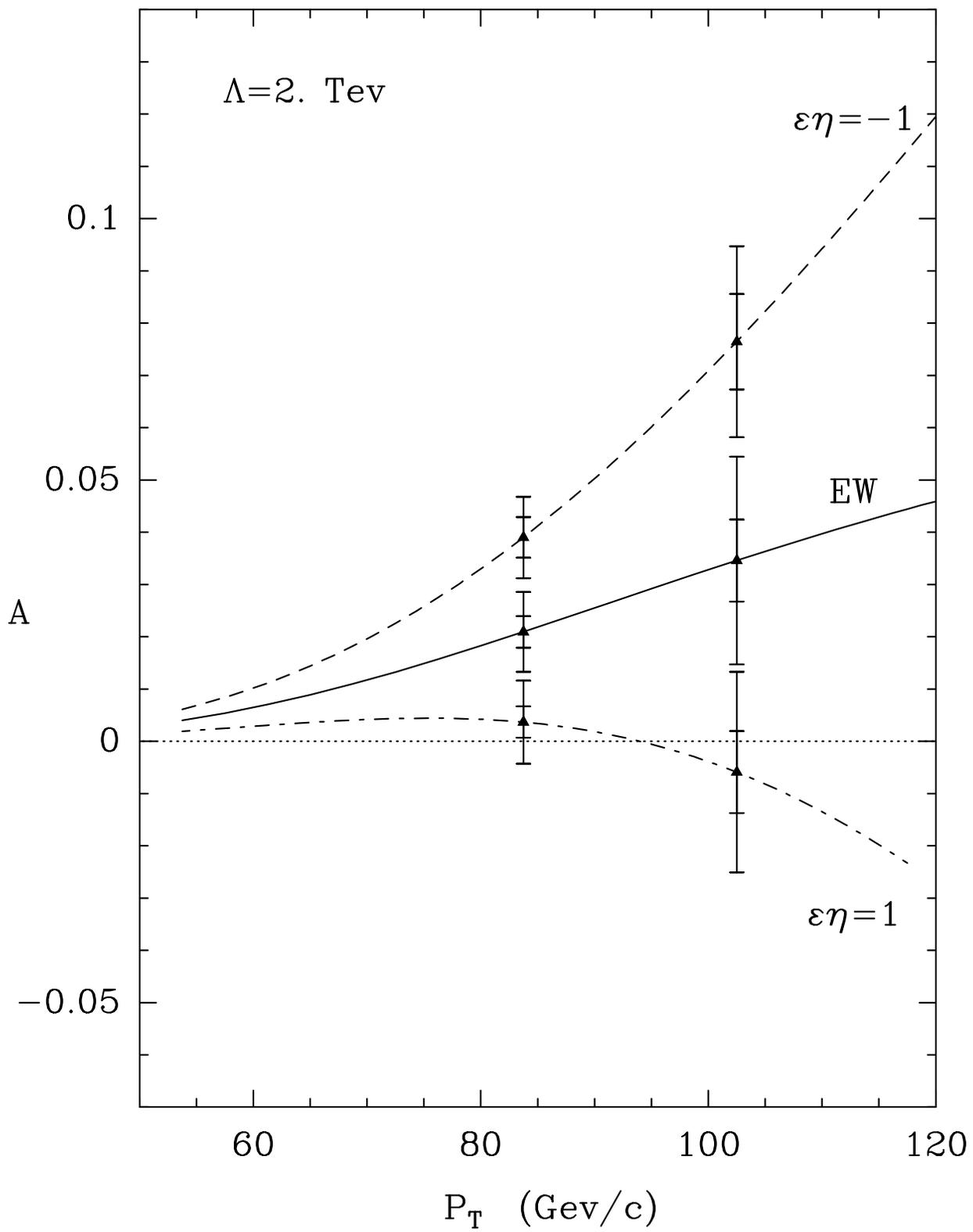
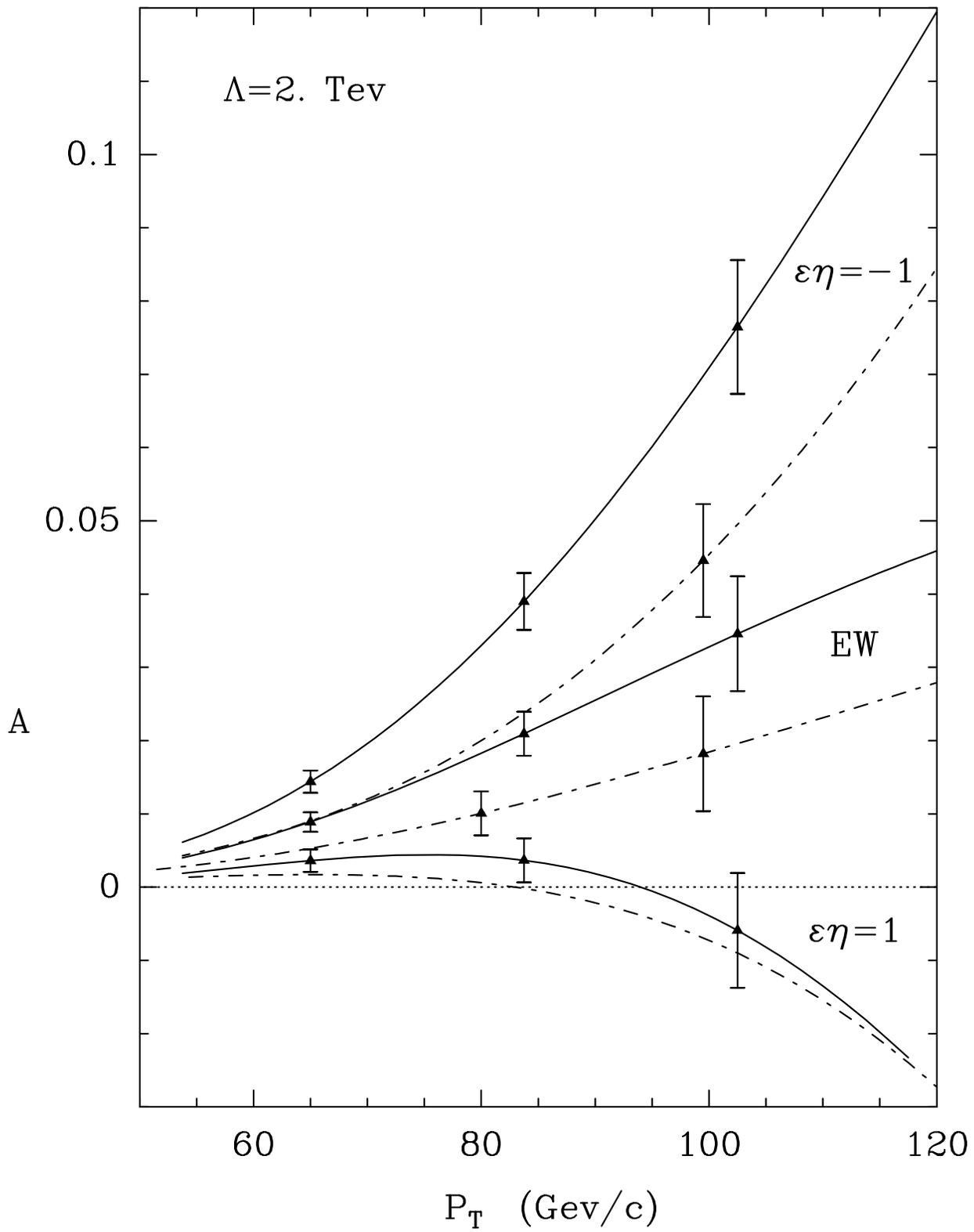
\end{document}